\documentstyle[11pt,epsf]{article}

\newcommand{\postscript}[2]
   {\setlength{\epsfxsize}{#2\hsize}
   \centerline{\epsfbox{#1}}}

\makeatletter\@addtoreset {equation}{section}\makeatother

\begin{document}

\begin{center}
{\Large {\bf Solitons in coupled Ablowitz-Ladik chains}}

\bigskip\bigskip

Boris A. Malomed

Department of Mathematics, University of Central Florida, Orlando, FL 32816,
USA, and Department of Interdisciplinary Studies, Faculty of Engineering,
Tel Aviv University, Tel Aviv 69978, Israel (permanent address); e-mail
malomed\@post.tau.ac.il

\bigskip

Jianke Yang

Department of Mathematics and Statistics, The University of Vermont,
Burlington, VT 05401, USA. email: jyang@emba.uvm.edu.

\bigskip

{\large Abstract}
\end{center}

A model of two coupled Ablowitz-Ladik (AL) lattices is introduced. While the
system as a whole is not integrable, it admits reduction to the integrable
AL model for symmetric states. Stability and evolution of symmetric solitons
are studied in detail analytically (by means of a variational approximation)
and numerically. It is found that there exists a finite interval of positive
values of the coupling constant in which the symmetric soliton is stable,
provided that its mass is below a threshold value. Evolution of the unstable
symmetric soliton is further studied by means of direct simulations. It is
found that the unstable soliton breaks up and decays into radiation, or
splits into two counter-propagating asymmetric solitons, or evolves into an
asymmetric pulse, depending on the coupling coefficient and the mass of the
initial soliton.

\newpage

\section{Introduction}

Dynamics of solitary waves (that we will refer to as ``solitons'', without
implying mathematical rigor) in nonlinear lattices is a vast field of
research, which is of great interest in its own right and finds important
physical applications. A majority of nonlinear discrete systems are
nonintegrable. However, there are paradigm models that are integrable by
means of the inverse scattering transform, most notably, the Toda lattice 
\cite{Toda} and Ablowitz-Ladik (AL) \cite{AL} chain. A nonintegrable
generalization of the latter model, in the form of its combination with the
discrete nonlinear Schr\"{o}dinger (NLS) equation,was also studied in detail 
\cite{Bishop,Bishop2}. Although the AL model finds fewer direct applications
than its discrete-NLS counterpart, it may describe, for instance,
ladder-structured lattices \cite{Vakh}.

The availability of exact soliton solutions in integrable models makes it
possible to study nontrivial dynamical effects in more complex systems that
can be built on the basis of integrable ones. In particular, a challenging
issue is to study solitons and their stability in two coupled systems, which
are integrable in isolation. A physically important example is a dual-core
optical fiber (the so-called directional coupler \cite{asymmetry}). While
each core is described by the integrable NLS equation, the coupled system is
not integrable, and it gives rise to a new effect: an obvious symmetric
soliton solution, with its energy equally split between the cores, becomes
unstable if the energy exceeds a certain critical value \cite{asymmetry}. As
a result of the onset of the instability, a pair of new solitons with a
broken symmetry emerge (they are mirror images to each other). Later,
similar bifurcations destabilizing symmetric solitons and replacing them by
asymmetric ones were found in a dual-core system with quadratic (rather than
cubic) nonlinearity \cite{Mak_chi^2}, and in a dual-core fiber with the
Bragg grating \cite{Mak_Bragg}.

A natural step is to introduce a system of two coupled AL chains and
investigate the stability and nonlinear evolution of its solitons. A system
of two AL chains with a coupling that does {\em not} admit reduction to the
usual AL model was introduced in Ref. \cite{Germans}, and solitary waves, as
well as moving breathers, were found in it. In this work, we focus on a
system of coupled AL chains that admits reduction to the integrable AL model
and seems, as a matter of fact, more natural. The system is 
\begin{equation}
i\dot{u}_{n}+(1+|u_{n}|^{2})\left[ (u_{n+1}+u_{n-1})+\epsilon
(v_{n+1}+v_{n-1})\right] =0,  \label{u}
\end{equation}
\begin{equation}
i\dot{v}_{n}+(1+|v_{n}|^{2})\left[ (v_{n+1}+v_{n-1})+\epsilon
(u_{n+1}+u_{n-1})\right] =0,  \label{v}
\end{equation}
where $u_{n}$ and $v_{n}$ are complex dynamical variables at the $n$-th site
of the chain, the overdot stands for the time derivative, and $\epsilon $ is
a real coupling constant. Under the symmetric reduction $u_{n}=v_{n}$, Eqs. 
(\ref{u}) and (\ref{v}) reduce to the AL model proper.

Results presented in this work demonstrate that the system of equations (\ref
{u}) and (\ref{v}) is definitely nonintegrable, unlike the AL model. The
present model conserves only two dynamical invariants, viz., the
Hamiltonian, 
\begin{equation}
H=-\sum_{n=-\infty }^{+\infty }\left[ \left( u_{n}u_{n-1}^{\ast
}+u_{n}^{\ast }u_{n-1}+v_{n}v_{n-1}^{\ast }+v_{n}^{\ast }v_{n-1}\right)
+\epsilon \left( u_{n}v_{n-1}^{\ast }+u_{n}^{\ast
}v_{n-1}+u_{n-1}v_{n}^{\ast }+u_{n-1}^{\ast }v_{n}\right) \right] ,
\label{H}
\end{equation}
and the total ``mass'', 
\begin{equation}
M=\sum_{n=-\infty }^{+\infty }\left[ \ln \left( 1+\left| u_{n}\right|
^{2}\right) +\ln \left( 1+\left| v_{n}\right| ^{2}\right) \right]  \label{N}
\end{equation}
(note that the masses corresponding to the $u$- and $v$-fields are not
conserved separately).

The symmetric reduction, $u_{n}(t)\equiv v_{n}(t)$, generates a solution to
Eqs. (\ref{u}) and (\ref{v}) which is tantamount to the well-known AL
soliton, 
\begin{equation}
u_{n}(t)=v_{n}(t)=\left( \sinh a\right) \,{\rm sech}\left( an+vt-\theta
_{0}\right) \exp \left[ -i\left( bn+\omega t-\phi _{0}\right) \right]
\label{exact}
\end{equation}
(in this paper, we only consider bright solitons). Here, $\theta _{0}$ and 
$\phi _{0}$ are position and phase constants, while parameters $a$ and $b$ determine the
soliton's velocity and frequency, 
\begin{equation}
v=2\left( 1+\epsilon \right) \sinh a\cdot \sin b,\,\,\omega =-2\left(
1+\epsilon \right) \cosh a\cdot \cos b  \label{vomega}
\end{equation}
[notice that the coupling parameter $\epsilon $ does not appear in the
expression (\ref{exact}) as it is absorbed into the time variable, which amounts to 
a rescaling of the velocity and 
frequency in Eq. (\ref{vomega})]. The value of the mass (\ref{N})
corresponding to the soliton (\ref{exact}) is 
\begin{equation}
M_{{\rm sol}}=4a.  \label{Nsol}
\end{equation}
In this work, we primarily focus on the stability of symmetric solitons (\ref
{exact}) with zero velocity, i.e., $b=\theta _{0}=0$.

The underlying equations (\ref{u}) and (\ref{v}) also allow 
the anti-symmetric reduction, $u_{n}(t)\equiv -v_{n}(t)$.  But there is no
necessity to treat this case separately, as it is tantamount to the
symmetric reduction with the change $\epsilon \rightarrow -\epsilon $. We
stress that, in this work, we consider both positive and negative values of 
$\epsilon $. In particular, the symmetric-soliton solution given by Eqs. (\ref
{exact}) and (\ref{vomega}) remains valid also in the case $1+\epsilon <0$.

The rest of the paper is organized as follows. In section 2, we put forward
an analytical approach to the stability problem, based on a variational
approximation (a review of the application of this technique to solitons was
recently given in Ref. \cite{review}). This approach, which we employ in its
simplest form, will produce partial information on the stability, as it
will be seen from comparison with results of numerical computation of
stability eigenvalues. The numerical eigenvalues are presented in section 3.
Finally, direct numerical simulations of the full system of Eqs. (\ref{u})
and (\ref{v}), showing nonlinear development of the instability, are
displayed in section 4.

\section{The variational approximation}

A feasible source of instability of the two-component soliton (\ref{exact})
is an eigenmode of small perturbations splitting the two components. To
accommodate this mode, we adopt the following {\it ansatz} for a perturbed
soliton, 
\begin{eqnarray}
u_{n}(t) &=&\left( \sinh a\right) \,{\rm sech}\left[ a\left( n-x(t)\right) %
\right] \exp \left[ +ic(t)n+i\phi (t)\right] ,  \label{uansatz} \\
v_{n}(t) &=&\left( \sinh a\right) \,{\rm sech}\left[ a\left( n+x(t)\right) %
\right] \exp \left[ -ic(t)n+i\phi (t)\right] ,  \label{vansatz}
\end{eqnarray}
where $2x(t)$ is a small time-dependent separation between centers of the $u$%
- and $v$-components, and $2c(t)$ is a dynamically conjugate variable, viz.,
a wavenumber difference between the components, while $\phi (t)$ is a common
time-dependent phase of both components. Note that this ansatz does not
include another possible perturbation mode, which may introduce a phase
difference between the two components of the soliton, its conjugate variable
being the amplitude difference between the components \cite{Kath}. We focus
on the restricted ansatz (\ref{uansatz}), (\ref{vansatz}) which accounts for
the position splitting, as its counterpart which also takes into
consideration the phase differences turns out to be cumbersome to calculate,
therefore it will not be pursued in this paper.

To apply the variational technique, we need a Lagrangian of the coupled
system (\ref{u}), (\ref{v}), which is 
\begin{equation}
L=\frac{i}{2}\sum_{n=-\infty }^{+\infty }\left[ \left( \dot{u}%
_{n}u_{n}^{\ast }-\dot{u}_{n}u_{n}^{\ast }\right) \frac{\ln \left( 1+\left|
u_{n}\right| ^{2}\right) }{\left| u_{n}\right| ^{2}}+\left( \dot{v}%
_{n}v_{n}^{\ast }-\dot{v}_{n}^{\ast }v_{n}\right) \frac{\ln \left( 1+\left|
v_{n}\right| ^{2}\right) }{\left| v_{n}\right| ^{2}}\right] -H,  \label{L}
\end{equation}
where $H$ is the Hamiltonian defined in Eq. (\ref{H}). An {\it effective
Lagrangian} can be calculated by substituting the ansatz (\ref{uansatz}), (%
\ref{vansatz}) into the Lagrangian (\ref{L}) and expanding it in powers of
the small separation parameters $x$ and $c$ up to quadratic terms, which are
necessary to generate perturbed equations of motion linear in $x$ and $c$.
Due to the obvious symmetry of the Lagrangian (\ref{L}), the time derivative
of $x$ will not appear in the result, while the time derivative $\dot{c}$ of 
$c$ may only appear linearly (being multiplied by $x$). Finally, making use
of some formulas for infinite sums related to the soliton's waveform
[involving ${\rm sech}\left( an\right) $], which were borrowed from Ref. 
\cite{Bishop2}), the effective Lagrangian is found in the following form 
\begin{equation}
L_{{\rm eff}}=-4\left[ a\dot{c}x+\left( \sinh a\right) \,\left(
c^{2}+Aa^{2}\epsilon \,x^{2}\right) \right] ,  \label{Leff}
\end{equation}
where $A$ is a positive constant defined as 
\begin{eqnarray}
A &\equiv &{\rm \sinh \,}a\sum_{n=-\infty }^{+\infty }\left\{ \frac{1}{2}
{\rm sech}\left( an\right) {\rm sech}\left( a\left( n-1\right) \right) \left[
{\rm sech}^{2}\left( an\right) +{\rm sech}^{2}\left( a\left( n-1\right)
\right) \right] \right.   \nonumber \\
&&\left. +{\rm sech}^{2}\left( an\right) {\rm sech}^{2}\left( a\left(
n-1\right) \right) \sinh \left( an\right) \sinh \left( a\left( n-1\right)
\right) \right\} -1\,.  \label{B}
\end{eqnarray}
In particular, $A=1\,$ in the continuum limit ($a\rightarrow 0$), and 
$A\approx 6\exp \left( -2a\right) $ in the ultradiscrete limit ($a\rightarrow
\infty $).

The effective Lagrangian (\ref{Leff}) immediately gives rise to an evolution
equation 
\begin{equation}
\ddot{x}+4\epsilon \left( \sinh ^{2}a\right) A\,x=0\,,  \label{evolution}
\end{equation}
which predicts that the symmetric soliton (\ref{exact}) is unstable against
the splitting perturbation mode if $\epsilon <0$. In this case, the
instability growth rate is 
\begin{equation}
\lambda =\left( 2\sinh a\right) \sqrt{-A\,\epsilon }\,,  \label{lambda}
\end{equation}
as predicted by Eq. (\ref{evolution}), i.e., by the variational
approximation.

\section{Numerical analysis of the instability of symmetric solitons}

A general stability analysis of the symmetric soliton (\ref{exact}) with 
$v=\theta _{0}=0$ assumes that a perturbed solution is taken as 
\begin{equation}
u_{n}=(U_{n}+\tilde{u}_{n})e^{-i\omega t},\,v_{n}=(U_{n}+\tilde{v}
_{n})e^{-i\omega t},\,U_{n}\equiv \left( \sinh a\right) \,\,{\rm sech}(an),
\label{perturbed}
\end{equation}
where the frequency $\omega $ is the same as defined in Eqs. (\ref{vomega})
(with $b=0$), and $\tilde{u}_{n},\tilde{v}_{n}$ are infinitesimal
perturbations. Numerically simulating equations (\ref{u}) and (\ref{v})
linearized about the unperturbed soliton solution, we have found that the
most unstable perturbation modes are always {\em anti-symmetric} ones, i.e.,
with $\tilde{u}_{n}=-\tilde{v}_{n}$ (which is not surprising, as the same is
true for a mode destabilizing symmetric solitons in all the previously
studied continuum models of the dual-core type \cite
{asymmetry,Mak_chi^2,Mak_Bragg}). Thus, we look for eigenmodes of the form 
\begin{equation}
\tilde{u}_{n}=-\tilde{v}_{n}=f_{n}\exp \left( \lambda t\right) +g_{n}^{\ast
}\exp \left( \lambda ^{\ast }t\right) ,  \label{eigen}
\end{equation}
where $\lambda $ is a (generally complex) instability growth rate.
Substituting Eqs. (\ref{perturbed}) and (\ref{eigen}) into Eqs. (\ref{u})
and (\ref{v}) and linearizing these equations, one eventually arrives at an
eigenvalue problem based on the equations 
\begin{equation}
\left( i\lambda +\omega \right) f_{n}+(1-\epsilon
)(1+|U_{n}|^{2})(f_{n+1}+f_{n-1})+(1+\epsilon
)U_{n}(U_{n+1}+U_{n-1})(f_{n}+g_{n})=0,  \label{eigenf}
\end{equation}
\begin{equation}
\left( -i\lambda +\omega \right) g_{n}+(1-\epsilon
)(1+|U_{n}|^{2})(g_{n+1}+g_{n-1})+(1+\epsilon
)U_{n}(U_{n+1}+U_{n-1})(g_{n}+f_{n})=0,  \label{eigeng}
\end{equation}
which are supplemented by the boundary conditions demanding that, for
discrete eigenmodes, the fields $f_{n}$ and $g_{n}$ vanish at $\left|
n\right| =\infty $.

The eigenvalue problem (\ref{eigenf}) and (\ref{eigeng}) was solved
numerically by the shooting method. Figure 1 presents unstable eigenvalues
as a function of the coupling constant $\epsilon $ for three different fixed
values of the soliton parameter, $a=0.8$, $0.9$, and $1$ [see Eqs. (\ref
{exact}) and (\ref{vomega})]. We have checked by simulating evolution
governed by the linearized equations that the unstable eigenvalues which are
presented in Fig. 1 are the most unstable ones. We suspect that these are
all the unstable eigenvalues in the above system (\ref{eigenf}) and (\ref
{eigeng}). Even if they are not, other eigenvalues are less unstable and
thus less significant.

Our numerical results for the eigenvalue problem show that, for each value
of $a$, the unstable eigenmodes at $\epsilon <0$ are odd, i.e., 
$f_{-n}=-f_{n}$ and $g_{-n}=-g_{n}$. When $\epsilon >0$, the eigenmodes are
even, i.e., $f_{-n}=f_{n}$ and $g_{-n}=g_{n}$. Furthermore, when negative 
$\epsilon $ has a small absolute value, the inspection of the unstable
eigenmode shows that it corresponds to a splitting of the two components
(appearance of separation between their centers) in the soliton 
(\ref{exact}). Recall that precisely this type of the perturbation 
mode was assumed in
the ansatz (\ref{uansatz}) and (\ref{vansatz}). On the other hand, when 
$\epsilon >0$ and small, the unstable eigenmode corresponds to a
phase-difference instability, which is not accounted for by the ansatz.

As is seen in Fig. 1, in the cases $a=0.9$ and $1.0$ the solitons (\ref
{exact}) are unstable for all values of $\epsilon $. The instability may 
be both non-oscillatory or oscillatory (corresponding to a real or complex
eigenvalue $\lambda $, respectively) in different intervals of $\epsilon $.
However, a stability window, $0.605<\epsilon <1.654$, is found at $a=0.8$.
More detailed computations show that the stability window opens up at a
critical value of the soliton parameter, $a_{{\rm cr}}\approx 0.881$, and
persists in the region $a<a_{{\rm cr}}$. At the point $a=a_{{\rm cr}}$, the
stability window appears (up to the accuracy of the numerical data) at the
value of the coupling constant $\epsilon _{{\rm cr}}=1$.

Inspection of Fig. 1 leads to the following general conclusions. First, all
symmetric solitons are unstable when $\epsilon <0$. In particular, near 
$\epsilon =-1$, the real part of the unstable eigenvalue is very small but
positive. Second, narrower solitons (with a larger mass, i.e., larger value
of $a$) are more unstable. Third, when the soliton mass is below the
threshold value $4a_{{\rm cr}}$, the soliton is stable inside a certain
positive-$\epsilon $ window. These features are somewhat similar to those in
the above-mentioned continuum models of the nonlinear-optical dual-core
systems with the linear coupling between the cores \cite
{asymmetry,Mak_chi^2,Mak_Bragg}. In those models too, symmetric solitons
become unstable if their mass exceeds a critical value, while the coupling
constant is positive, and they are never stable if the coupling constant is
negative. However, the stability window observed in Fig. 1 when 
$a<a_{{\rm cr}}$ is different from stability regions 
in the continuum dual-core models,
as the latter stability regions have no right endpoint (they are
semi-infinite).

It is natural to compare the numerically found unstable eigenvalues with the
one (\ref{lambda}) that was predicted by the variational approximation in
the previous section. This comparison is meaningful only when $\epsilon <0$
and $|\epsilon |\ll 1$, as the actual unstable mode corresponds to the
variational ansatz (\ref{uansatz}) and (\ref{vansatz}) only in this case.
For $a=0.8$, the comparison is displayed in Fig. 2. It is not surprising
that the deviation between the analytical and numerical results is large for
large values of $\left| \epsilon \right| $, when the coupling term can make
the perturbed soliton strongly different from the ansatz (\ref{uansatz}), 
(\ref{vansatz}). For small $|\epsilon |$, the analytical value is close to
the numerical one; the remaining difference might be due to the fact that,
if the two components of the soliton are separated by some distance, the
symmetry of each component relative to its center is broken by the coupling
term, which is not taken into regard by the ansatz.

\section{Direct simulations of the instability development}

In order to understand the evolution of unstable symmetric solitons, direct
numerical simulations of the full nonlinear equations (\ref{u}) and (\ref{v})
were performed. To this end, the initial configuration was taken in the
form corresponding to the ansatz (\ref{u}), (\ref{v}), with an extra
perturbation parameter, namely, a phase difference $2\psi $ between the two
components of the soliton: 
\begin{eqnarray}
u_{n}(t) &=&\left( \sinh a\right) \,\,{\rm sech}\left( a\left(
n-x_{0}\right) \right) \exp \left( -ic_{0}n+i\psi _{0}\right) ,
\label{upert} \\
v_{n}(t) &=&\left( \sinh a\right) \,{\rm sech}\left( a\left( n+x_{0}\right)
\right) \exp \left( +ic_{0}n-i\psi _{0}\right) .  \label{vpert}
\end{eqnarray}

Different typical outcomes of the instability development can be illustrated
by a set of contour-plot pictures pertaining to several characteristic
values of $\epsilon $, while all the other parameters are fixed. In Figs. 3
through 7, we display the pictures for a representative case $a=0.8$, 
$x_{0}=c_{0}=\psi _{0}=0.01$. In fact, exact initial values of the small
perturbations are not important, while the value of the soliton parameter $a$
is a significant one. In each figure, contours in the left and right panels
represent the evolution of $\left| u_{n}\right| $ and $\left| v_{n}\right|$, 
respectively. All the contours start at the level $0.05$ and increase with
an increment of $0.2$.

First of all, Fig. 3 shows that if $\epsilon $ belongs to the stability
interval, see the upper panel in Fig. 1 (in Fig. 3, $\epsilon =1$), the
initial perturbation indeed does not trigger any instability. Next, Fig. 4
shows that, in the case $\epsilon =-7$, which corresponds to strong
instability, the soliton gets completely destroyed, decaying into radiation.
Decreasing the absolute value of negative $\epsilon $, i.e., proceeding to
weaker instability, according to Fig. 1, we observed a trend of splitting of
the original unstable symmetric soliton into two {\em moving} ones, each
being a two-component pulse. An example of that is shown in Fig. 5 for 
$\epsilon =-0.08$. It is noteworthy that the symmetry of each secondary
soliton is strongly broken (the amplitude of one component is definitely
larger than that of the other), but they are (at least, approximately)
mirror images to each other, so that the global symmetry is conserved.

In the case when the symmetric soliton is unstable at positive $\epsilon$,
the character of the instability is quite different from that described
above for $\epsilon <0$. In this case, the instability is non-oscillatory.
In Fig. 6, which pertains to $\epsilon =0.3$, a noteworthy result is the
formation of a soliton with a strongly broken symmetry and (nearly) periodic
internal oscillations, which is accompanied by emission of small amounts
of radiation. This instability-induced spontaneous symmetry
breaking resembles what is known in the above-mentioned continuum models of
the dual-core type \cite{asymmetry,Mak_chi^2,Mak_Bragg}. Of course, in all
the cases when spontaneous symmetry breaking is observed, its sign (e.g.,
the difference between the left and right panels in Fig. 6) is determined 
by a random initial perturbation.

Finally, Fig. 7 shows the situation in the case $\epsilon =2$, which lies
beyond the right border of the stability interval in Fig. 1 (recall that no
such border occurs in the continuum dual-core models). As is seen, in this
case the instability also breaks the symmetry of the soliton (in the
beginning). However, an unusual feature here is that the centers of both
components oscillate, essentially, in-phase. Unlike what was seen in all the
other figures, spontaneous onset of such in-phase oscillations seems to
violate the momentum conservation; however, one should keep in mind that
lattice systems conserve no momentum, in view of the lack of the continuous
translational invariance in them. Nevertheless, the appearance of such a
dynamical state is a remarkable fact which may deserve further investigation.

\section{Conclusion}

In this work, we have introduced a model of two coupled Ablowitz-Ladik
chains. While the system as a whole is not integrable, it admits reduction
to the integrable AL model for symmetric states. We have studied the
stability and nonlinear evolution of stationary symmetric solitons in
detail. Both the analytical consideration, based on the variational
approximation, and numerical computation of the instability eigenvalues have
demonstrated that the soliton may be unstable. Numerical results also show
that, provided that the soliton's mass is below a critical value, there
exists a finite interval of positive values of the coupling constant 
$\epsilon $ in which the symmetric soliton is stable. Comparison of the
approximate analytical and exact numerical unstable eigenvalues shows that
the agreement is reasonable for small negative values of the coupling
constant. Evolution of the unstable symmetric soliton was further studied by
means of direct simulations. It was found that the unstable soliton can
decay into radiation, or split in two counter-propagating asymmetric
solitons, or evolve into an asymmetric pulse, depending on the values of the
coupling coefficient $\epsilon $ and the soliton mass.

\section{Acknowledgements}

B.A.M. appreciates hospitality of the Department of Mathematics and
Statistics at the University of Vermont, and of the Department of
Mathematics at the University of Central Florida. J.Y. acknowledges partial
financial support from the Air Force Office of Scientific Research and the
National Science Foundation.

\newpage

\begin{figure}[tbp]
\begin{center}
\parbox{13cm}{\postscript{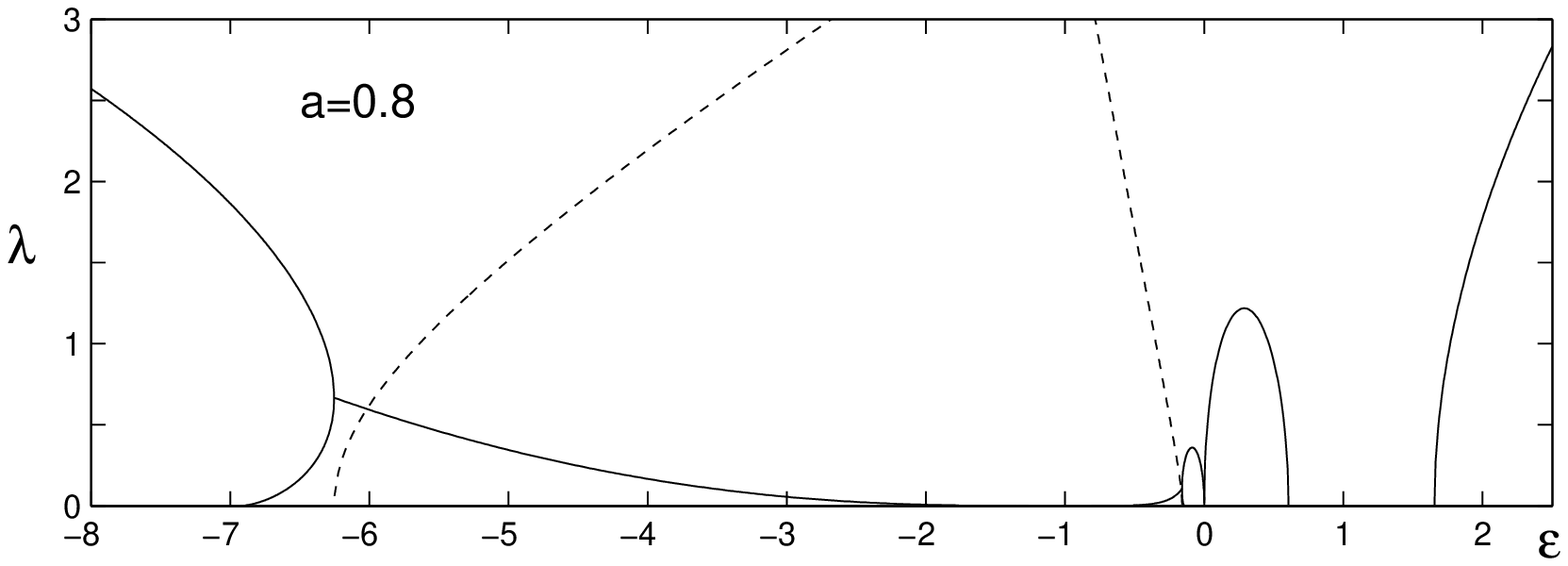}{1.0}}
\par
\parbox{13cm}{\postscript{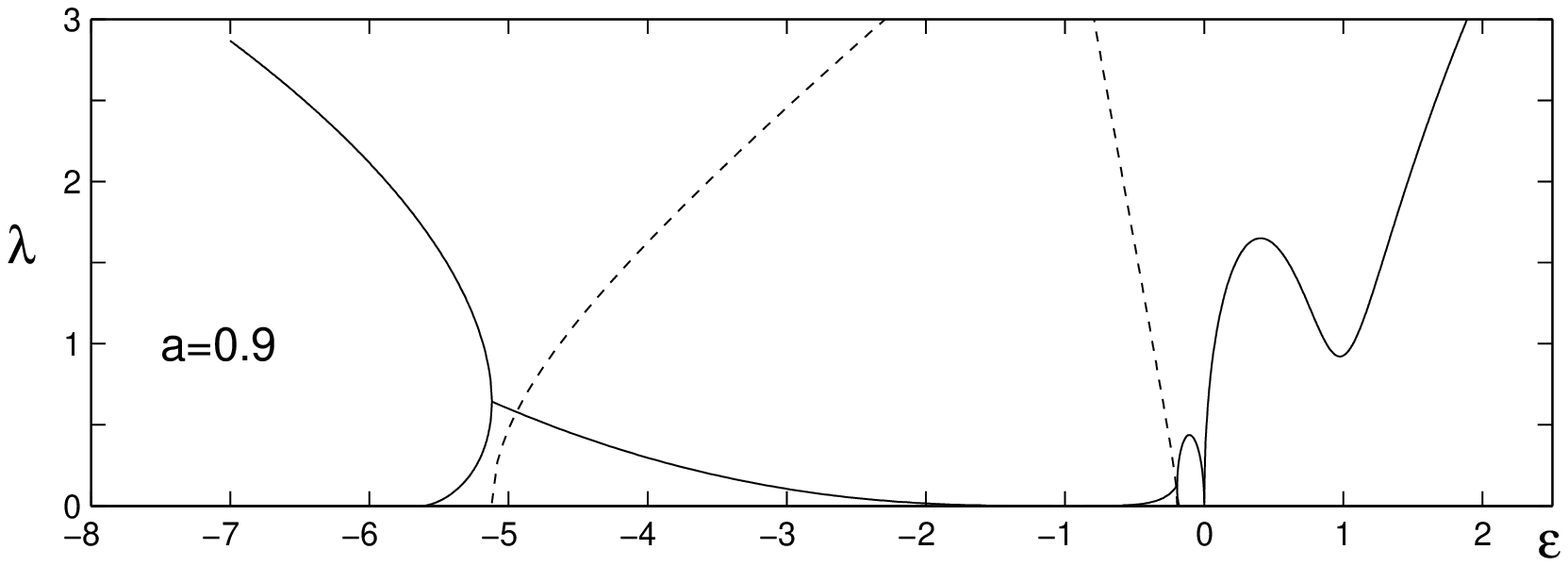}{1.0}}
\par
\parbox{13cm}{\postscript{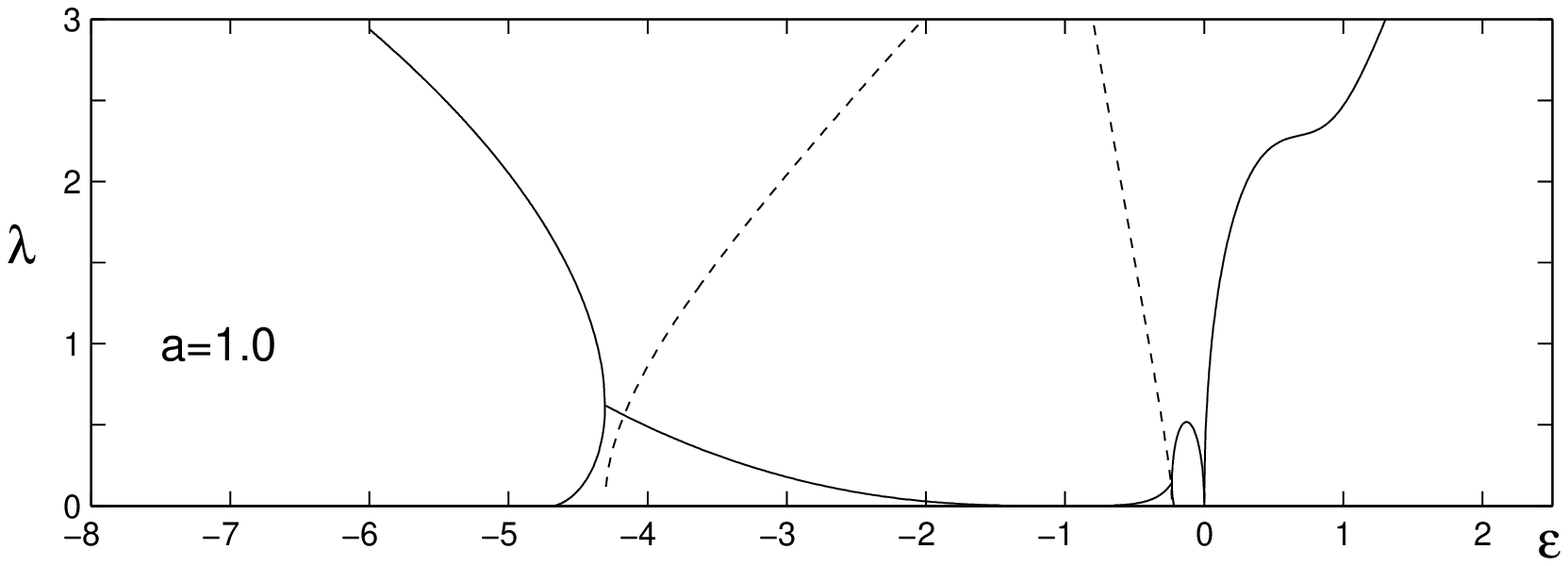}{1.0}}
\end{center}
\caption{The numerically found unstable eigenvalue $\protect\lambda $ versus
the coupling coefficient $\epsilon$ for
the symmetric stationary soliton at different fixed values of the soliton
parameter $a$ [see Eqs. (\ref{exact}) and (\ref{vomega})]. The solid and
dashed curves show, respectively, the real and imaginary parts of $\protect
\lambda$. The imaginary part is not shown inside the stability window in the
upper panel, where $\protect\lambda$ is pure imaginary. In the interval
around the point $\protect\epsilon = -1$, the real part of $\protect\lambda$
is very small but finite in all three panels.}
\end{figure}

\begin{figure}[tbp]
\begin{center}
\parbox{8cm}{\postscript{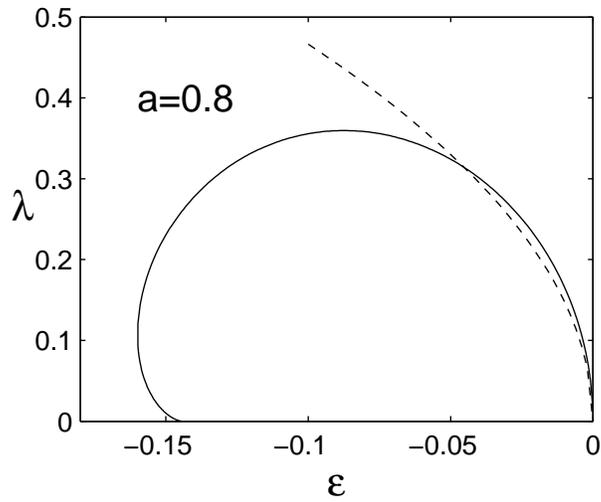}{1.0}}
\end{center}
\caption{The instability growth rate predicted by the variational
approximation, see Eq. (\ref{lambda}) (the dashed curve), vs. its
numerically computed counterpart (the solid curve).}
\end{figure}

\begin{figure}[tbp]
\begin{center}
\parbox{14cm}{\postscript{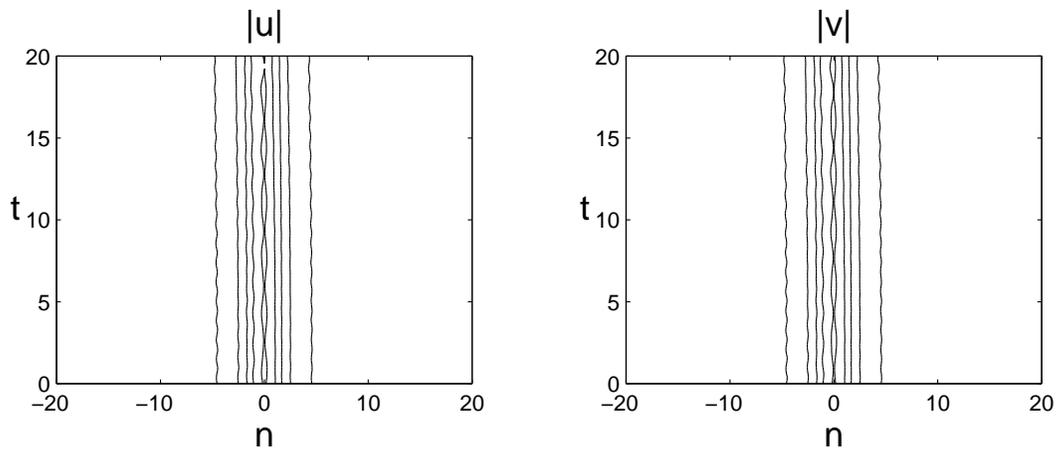}{1.0}}
\end{center}
\caption{Evolution of the initially perturbed symmetric soliton in the case 
$a=0.8$, $\protect\epsilon =1$, when the soliton is stable. In this figure
and below, all contours start at the level $0.05$ and increase with an
increment of $0.2$.}
\end{figure}

\begin{figure}[tbp]
\begin{center}
\parbox{14cm}{\postscript{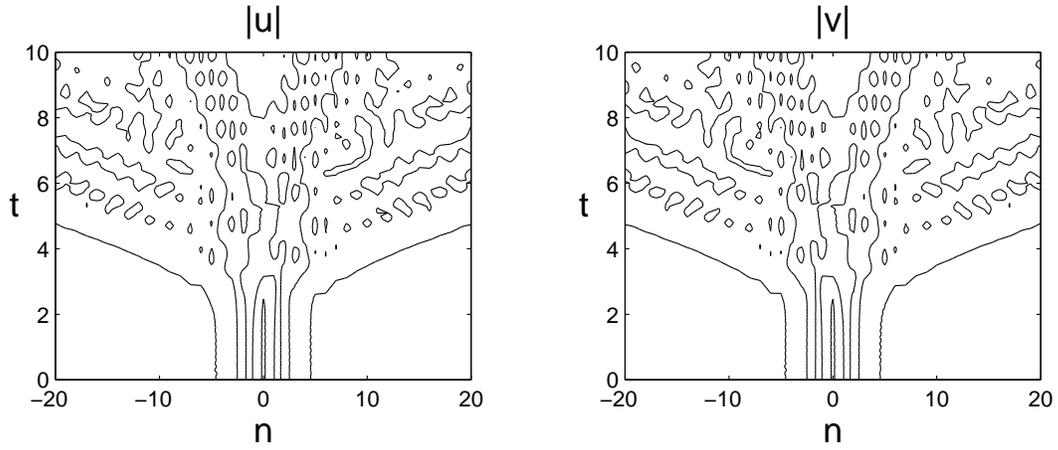}{1.0}}
\end{center}
\caption{Evolution of an unstable symmetric soliton in the case $a=0.8$, 
$\protect\epsilon =-7$.}
\end{figure}

\begin{figure}[tbp]
\begin{center}
\parbox{14cm}{\postscript{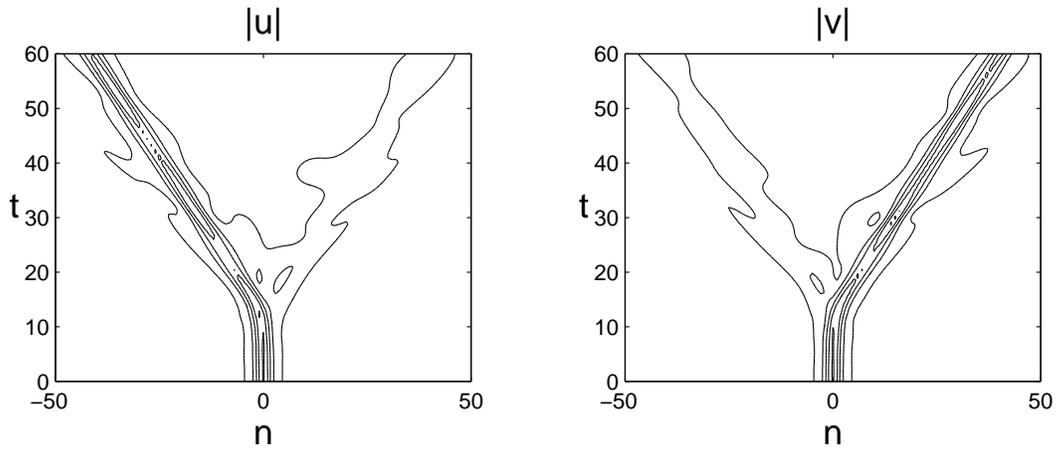}{1.0}}
\end{center}
\caption{Splitting of an unstable symmetric soliton in the case $a=0.8$, 
$\protect\epsilon =-0.08$.}
\end{figure}

\begin{figure}[tbp]
\begin{center}
\parbox{14cm}{\postscript{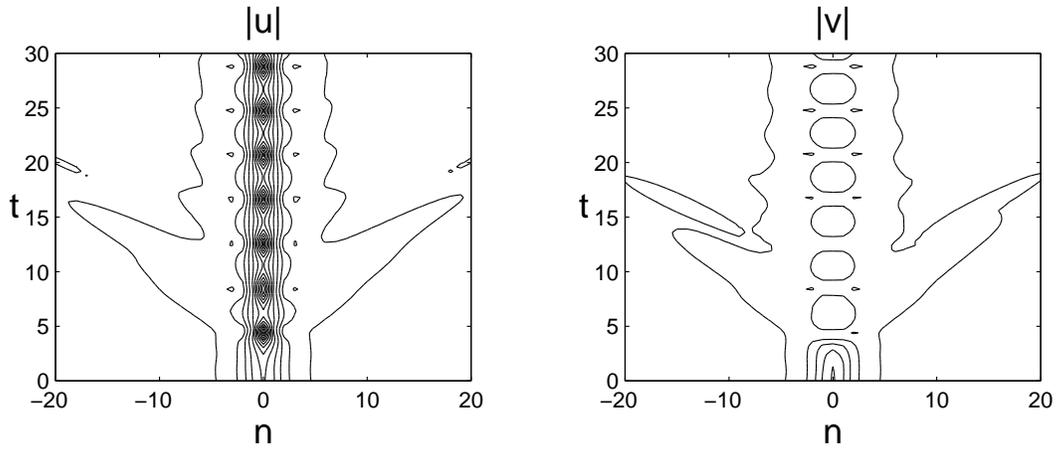}{1.0}}
\end{center}
\caption{Conversion of an unstable symmetric soliton into a stable
asymmetric one in the case $a=0.8$, $\protect\epsilon =0.3$.}
\end{figure}

\begin{figure}[tbp]
\begin{center}
\parbox{14cm}{\postscript{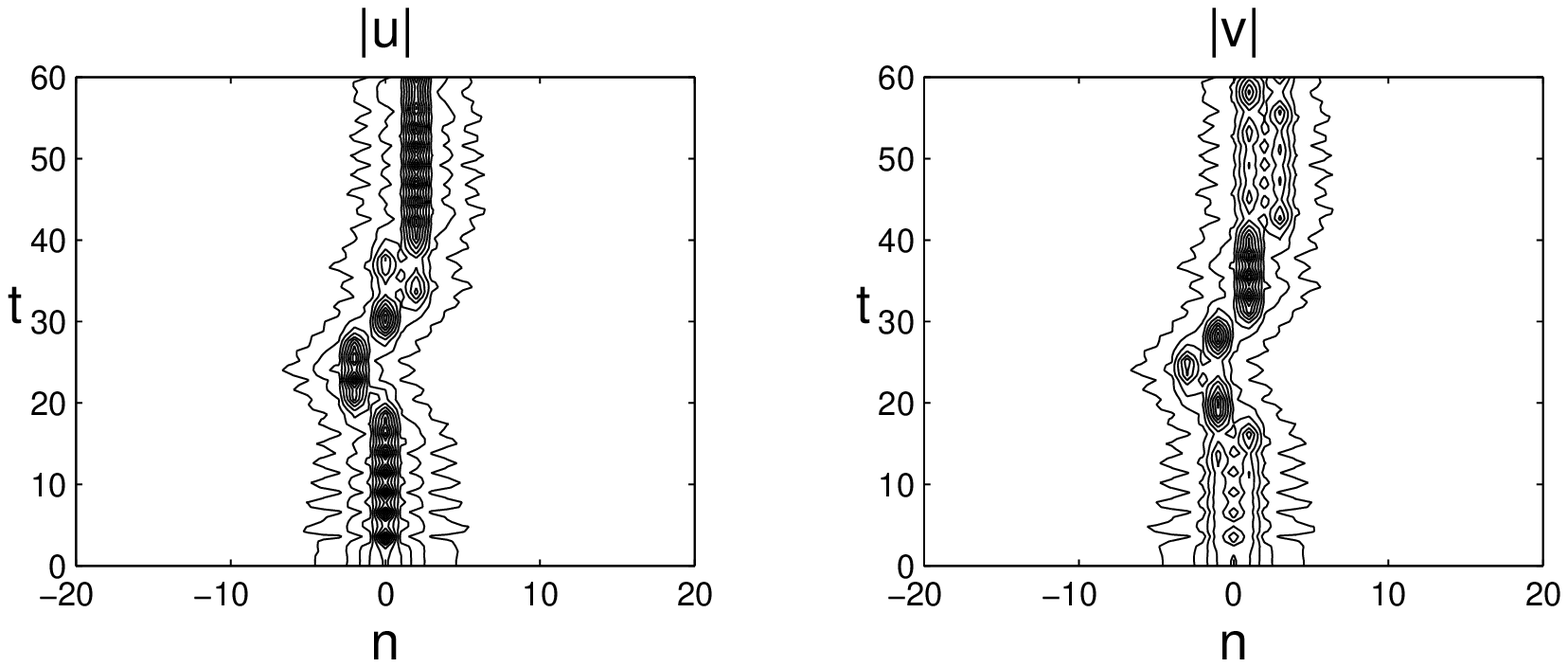}{1.0}}
\end{center}
\caption{Evolution of an unstable soliton in the case $a=0.8$, $\protect
\epsilon =2$.}
\end{figure}


\newpage

\begin{thebibliography}{99}
\bibitem{Toda}  M. Toda, Phys. Rep. {\bf 18}, 1 (1973).

\bibitem{AL}  M.J. Ablowitz and J.F. Ladik, J. Math. Phys. {\bf 17}, 1011
(1976).

\bibitem{Bishop}  D. Cai, A.R. Bishop, and N. Gr\o nbech-Jensen, Phys. Rev.
E {\bf 53}, 4131 (1996).

\bibitem{Bishop2}  D. Cai, A.R. Bishop, and N. Gr\o nbech-Jensen, Phys. Rev.
E {\bf 56}, 7246 (1997).

\bibitem{Vakh}  O.O. Vakhnenko and V.O. Vakhnenko, Phys. Lett. A {\bf 196},
307 (1995); O.O. Vakhnenko, Phys. Rev. E {\bf 64}, 067601 (2001).

\bibitem{asymmetry}  N. Akhmediev, and A. Ankiewicz, Phys. Rev. Lett. {\bf
70 }, 2395 (1993); P.L. Chu, B.A. Malomed, and G.D. Peng, J. Opt. Soc. Am. B 
{\bf 10}, 1379 (1993).

\bibitem{Mak_chi^2}  W. Mak, B.A. Malomed, and P.L. Chu, Phys. Rev. E 55,
6134 (1997).

\bibitem{Mak_Bragg}  W. Mak, B.A. Malomed, and P.L. Chu, J. Opt. Soc. Am. B 
{\bf 15}, 1685 (1998).

\bibitem{Germans}  A. B\"{u}low, D. Hennig, and H. Gabriel, Phys. Rev. E 
{\bf 59}, 2380 (1999).

\bibitem{review}  B.A. Malomed, Progress in Optics {\bf 43}, 69 (2002).

\bibitem{Kath}  T. Ueda and W.L. Kath, Phys. Rev. A {\bf 42}, 563 (1990).
\end{thebibliography}
\end{document}